\documentclass[smallextended]{svjour3}

\usepackage[utf8]{inputenc}
\usepackage[english]{babel}
\usepackage{mathrsfs}
\usepackage{bm}
\usepackage{graphicx}
\usepackage{epstopdf}

\journalname{Eur. Phys. J. B}

\begin{document}

\title{Probing magneto-elastic phenomena through an effective spin-bath coupling model}
\titlerunning{Probing magneto-elastic phenomena through an effective spin-bath coupling}

\author{Thomas Nussle \and Pascal Thibaudeau \and Stam Nicolis}
\institute{T.Nussle \email{thomas.nussle@cea.fr} \and P.Thibaudeau \email{pascal.thibaudeau@cea.fr} \at CEA DAM/Le Ripault, BP 16, F-37260, Monts, FRANCE
\and T.Nussle \email{thomas.nussle@cea.fr}  \and S.Nicolis \email{stam.nicolis@lmpt.univ-tours.fr} \at Institut Denis Poisson, UMR CNRS 7013, Université de Tours, Université d’Orléans,
Parc de Grandmont 37200, Tours, FRANCE}

\maketitle

\begin{abstract}
	A phenomenological model is constructed, that captures the effects of coupling magnetic and elastic degrees of freedom, in the presence of external, stochastic perturbations, in terms of the interaction of magnetic moments with a bath, whose individual degrees of freedom cannot be resolved and only their mesoscopic properties are relevant.
	In the present work, the consequences of identifying the effects of dissipation as resulting from interactions with a bath of spins are explored, in addition to elastic, degrees of freedom.
	The corresponding stochastic differential equations are solved numerically and the moments of the magnetization are computed.
	The stochastic equations implicitly define a measure on the space of spin configurations, whose moments at equal times satisfy a hierarchy of deterministic, ordinary differential equations.
	Closure assumptions are used to truncate the hierarchy and the same moments are computed.
	We focus on the advantages and problems that each approach presents, for the approach to equilibrium and, in particular, the emergence of longitudinal damping.
\end{abstract}

\section{Introduction}\label{Introduction}
With the continuous reduction in size of magnetic systems, thermally induced fluctuations of the magnetization are responsible for limiting the signal-to-noise ratio of such devices \cite{smith_white-noise_2001}.
How to describe the physical degrees of freedom that give rise to the magnetic fluctuations on the nanometer scale represents a difficult challenge \cite{bacher_monitoring_2002}.
Besides, noise sources contain valuable information about the system itself \cite{crooker_spectroscopy_2004}.
This motivates developing classes of models for the microscopic degrees of freedom that can describe the noise--and, equally importantly--how the distinction of these from the ``physical" degrees of freedom that can be identified with the magnetization, can be understood as a choice of coordinates in an enlarged phase space, that describes a consistently closed system.
The identities between generalized susceptibilities that are the consequence of this symmetry can be checked by numerical simulations and can lead to protocols for experiments, that have become of practical relevance \cite{degen_quantum_2017,forster_hydrodynamic_1994}.

A particular challenge in this field is understanding how damping of the magnetic fluctuations, parallel to the applied field (these are called  ``longitudinal"), first of all, can emerge and, next can be related to the damping of the fluctuations that are transverse to the applied field.
While there have been many attempts in the literature~\cite{kambersky_ferromagnetic_1976,garanin_fokker-planck_1997}, a general, coherent, picture is, still lacking.
One objective of the present study is to pursue the approach set out in~\cite{thibaudeau_nambu_2017} and provide more tangible evidence of how longitudinal damping can emerge and be related to transverse damping.

The general idea for describing magnetization fluctuations is to start with a stochastic generalization of the Landau-Lifshitz-Gilbert equation of motion for a collection of spins \cite{brown_thermal_1963,kubo_brownian_1970,brown_thermal_1979,coffey_thermal_2012}.
The physical origin of such a vector stochastic noise is not generally specified and only its statistical properties, defined by its moments are assumed known \cite{bertotti_nonlinear_2009}.
One way of setting up a corresponding Lagrangian formalism is through the Caldeira-Leggett model \cite{caldeira_quantum_1983}, which represents a semi-empirical way to deal with the dynamics of a system in equilibrium with  an environment described as a collection of commuting, harmonic, oscillators.

Depending on the constraints imposed by the classical equations of motion of the system variables, such a model leads to a Hamiltonian model  in the canonical ensemble, that mimics stochastic character of the realistic physical phenomenon and remains analytically tractable.

This approach has, also, been widely used in the context of open quantum systems~\cite{zwanzig_nonlinear_1973}, when the fluctuations are taken to be quantum.

Surprisingly, it took more than 20 years to see such ideas applied to the dynamics of a quantum magnet, coupled to  quantum and thermal baths, that are assumed to be described by elastic modes~\cite{rebei_fluctuations_2003}.
In such an approach, an undamped and non-fluctuating spin is coupled to a reservoir of scalar degrees of freedom, whose dynamics is that of harmonic oscillators, and whose properties define the reservoir.
The detailed description of the magnetoelastic coupling was not considered at that time but was investigated later \cite{rossi_dynamics_2005} by introducing a strain tensor, that was defined through its expansion in terms of bosonic, collective harmonic modes.

In a previous work~\cite{nussle_coupling_2019} we studied the effect of the coupling of spin degrees of freedom to elastic degrees of freedom, that resolved a corresponding bath, on the switching properties of the magnetization, described through the spins.

In the present work, we explore the effects of the polarization that a bath can have, when it is described as a spin, itself.

Once more, the phase space of the considered spin is, first, enhanced by the  additional degrees of freedom that couple realistically with it.
These extra variables resolve the effective stochastic bath. 
Such a bath has, in fact, been the focus of recent studies, that are reviewed in~\cite{prokofev_theory_2000}.

The consequences of identifying the effects of dissipation as resulting from interactions with a spin bath, rather than with elastic degrees of freedom are explored in an approximation that involves solving the corresponding stochastic differential equations numerically. 

In summary we describe the spin bath as a ``heavy" spin, coupled to a ``light" spin, that describes the dynamical degrees of freedom of the magnet. We then use a ``vanishing-cumulants" approximation to factorize the mixed moments of ``heavy" and ``light" spins, in a way that can lead to the emergence of longitudinal damping for the ``light" spin.
This strategy has been largely adopted in different contexts \cite{casalbuoni_effective_1993}.
Once we obtain a closed--effective--hierarchy for the moments of those distributions, we evaluate them numerically, while simultaneously integrating the stochastic equation with a homemade symplectic/geometric integrator so as to check our hypothesis.
We then check for the emergence of longitudinal damping, especially on the light spin, and study a few cases to see how this coupling might be able to display magneto-elastic behavior.

We then open a discussion for extending this model with a bath of fluctuating tensor degrees of freedom, and how this extension could enrich our model.

\section{Review of the system-plus-bath approach}\label{system+bath}

Our starting point is a spin ${\bm s}$, that performs a precession motion about an external field labelled ${\bm\omega}$.
This field is assumed to receive feedback from the state of the spin itself and to be subject to additional degrees of freedom, $\{x_i\}$, $i=1,2,\ldots,N$, whose dynamics is taken to be defined by ordinary differential equations:
\begin{equation}
\left\{
\begin{array}{rcl}
\dot{\bm{s}} &=& \bm{\omega}(\bm{s},x_i)\times\bm{s} \\
\dot{x}_i &=& f_i({\bm s},{x_i})
\end{array}
\right.
\label{eoms}
\end{equation}
It is, further, assumed that the effect of the additional degrees of freedom, $\{x_i\}$ is to drive the system to some attractor,
 by shrinking its phase-space volume over time, i.e.
\begin{equation}
\frac{\partial\dot{\bm s}}{\partial{\bm s}}+\frac{\partial{\dot x}_i}{\partial{x}_i}<0
\end{equation}
An explicit example of this scenario was, indeed, presented some years ago~\cite{tranchida_quantum_2015} where it was found that the interplay between Bloch--Bloembergen dissipation and an external torque implies that the Landau--Lifshitz--Bloch equations, augmented with the external torque, are equivalent to the equations for the Lorenz attractor; thereby establishing that this type of dissipation describes deterministic chaos and that the equilibrium measure is, in fact, fractal.
The existence of such a chaotic behavior in the context of magnetization dynamics has been debated and is still the subject to some controversial issues \cite{bertotti_nonlinear_2001,alvarez_quasiperiodicity_2000}.

It should be stressed that the existence of such an equilibrium measure is quite non--trivial, because the motion of precession does not derive from a scalar potential; therefore the dissipation that can be applied to it does not lead to a point attractor, as is the case for usual potential motion.
This fact has not received the attention it deserves.

So what we wish to explore is how the scenario of ref.~\cite{tranchida_quantum_2015} can be generalized, when the dissipative dynamics is not necessarily of Bloch--Bloembergen form.

Since equations~(\ref{eoms}) are deterministic, a given set of initial
conditions 
allows to integrate them formally and thus
define a Liouville operator ${\cal{L}}$, such that for any time $t$,
\begin{equation}
({\bm s}(t),x_i(t))=\exp({t\cal{L}})({\bm s}(0),x_i(0)).
\end{equation}
${\cal L}$ can be written as the sum of two operators ${\cal L}_s$ and ${\cal L}_x$, that describe the solutions of the corresponding equation of motion, assuming the other variable as constant.
Unfortunately, the exponential of the sum of these operators is not the product of the exponential of these operators in general \cite{blanes_magnus_2009}, which is particularly true for spins \cite{krech_fast_1998}.
Such a formal procedure does not lead to particularly transparent expressions for the pair $({\bm s}(t),x_i(t))$, in general, but very
good approximation schemes for the solutions can be efficiently computed by noting the geometric nature of such an integration process
\cite{hairer_geometric_2006}.

However this approach does break down for deterministic chaotic systems, where its technical failure is due to the fact that the attractor is of measure zero in the space of configurations.
In this context, a suitable splitting of such a closed dynamical system amounts to enlarging the underlying phase-space of dynamical variables, $\bm{s}$ by including the variables $\{x\}$.

Such a procedure was studied in detail for spin degrees of freedom and allows a way to define a thermostat for the atomic spin dynamics from controlled demons \cite{thibaudeau_thermostatting_2012}.
This approach, therefore, leads to a consistent thermodynamics, by identifying the chaotic contribution as ``molecular chaos" \`a la Boltzmann, rather than as ``deterministic chaos" \`a la Lorenz \cite{schuster_deterministic_2006}.

We wish, therefore, to explore the possibility of identifying the degrees of freedom $\{x\}$ with the components of a heavy spin, $\bm{S}$.
In order to get the physical insights from that, some correspondence between the elastic properties of the underlying solid and an effective heavy spin that mimic them can be seen manifest, when the heavy spin component ${\bm S}$ is coupled via the magneto-elastic coupling tensor $B$ such as
the instantaneous Cauchy strain tensor $e_{ij}$ is \cite{du_tremolet_de_lacheisserie_magnetostriction:_1993}
\begin{equation}
		e_{ij}=-\frac{1}{2}S_{ijmn}B_{mnkl}S_kS_l,
\end{equation}
where $S_{ijmn}$ are the components of the elastic compliance tensor.
Moreover, it has been demonstrated that a fully coupled dynamical system,
between the mechanical strain and the magnetic moment can be established
through a Lagrangian description \cite{nussle_coupling_2019}.
However, a stochastic variant of that approach has to be developed.
Considering ${\bm S}$ is a convenient shortcut to allow the immediate
use of the well-developed machinery of the stochastic spin dynamics \cite{skubic_method_2008}, including the possibility to experimentally
probe its dynamics via a spectroscopy of spontaneous spin noise \cite{crooker_spectroscopy_2004}.
This is the subject of the next section.

\section{The heavy spin as a bath}\label{heavySpin}

This relies on resolving several conceptual issues. 
From the collection of a large quantity of spins, we isolate a single one and call it the light spin ${\bm s}$.
This light spin, with no intrinsic damping, is coupled by an exchange interaction to a heavy macrospin ${\bm S}$.
Such a heavy spin is coupled, in turn, to a random field, ${\bm\eta}$, and interacts, with the light spin through an exchange interaction according to the Landau--Lifshitz--Gilbert equation.
Therefore the full system is described by the following equations:
\begin{equation}
\label{HeavyLight}
\left\{
\begin{array}{rcl}
\dot{\bm{s}} &=& \bm{\omega}(\bm{S})\times\bm{s} \\
\dot{\bm{S}} &=& \left[\bm{\Omega}(\dot{\bm S},{\bm s})+\bm{\eta}\right]\times\bm{S}
\end{array}
\right.
\end{equation}
To be specific, we shall take $\bm{\eta}$ as an Ornstein-Uhlenbeck process defined by
\begin{equation}
\left\{
\begin{array}{rcl}
\langle{\bm\eta}(t)\rangle&=&{\bm 0}\\
\langle\eta_i(t)\eta_j(t')\rangle&=&\displaystyle{\frac{D}{\tau}\exp\left(\frac{-|t-t'|}{\tau}\right)\delta_{ij}}
\end{array}
\right.
\end{equation}
where $D$ is the noise amplitude and $\tau$ is the--finite--time of response of the bath.
Moreover, since the noise distribution is assumed to be described by a Gaussian process, all the higher order moments are given by combinations of the first and second moments only, according to Wick's theorem~\cite{van_kampen_stochastic_1976}.
This form of noise, which is assumed to describe the elastic medium, in which the spins are immersed, will prove particularly useful for technical reasons discussed below. 

The noise is assumed to affect only the heavy spin ${\bm S}$, but through the coupling, both spins acquire a non--trivial probability distribution, whose moments we wish to calculate.

We wish to focus, in particular, on the moments of the light spin distribution and to do so we shall compute appropriate averages over the noise.

As an additional simplification, we will start by focusing on external fields given by $\bm{\omega}({\bm S})\equiv\bm{\omega}+J{\bm S}$ and $\bm{\Omega}(\dot{\bm S},{\bm s})\equiv\bm{\Omega}-\alpha\dot{\bm S}+J{\bm s}$, where $\bm{\omega}$ and ${\bm\Omega}$
are constant fields, $\alpha$ is a damping coefficient and $J$ is an exchange parameter.

Under these assumptions eqs.~(\ref{HeavyLight}) take the form
\begin{equation}
\label{HLextfields}
\left\{
\begin{array}{rcl}
\displaystyle
\dot{\bm{s}} &=& (\bm{\omega}+J\bm{S})\times\bm{s} \\
\displaystyle
\dot{\bm{S}} &=& (\bm{\Omega}-\alpha\dot{\bm S}+J{\bm s}+\bm{\eta})\times\bm{S}
\end{array}
\right.
\end{equation}
It should be stressed that this way of introducing the noise implies that the norms of the light and heavy spins are, stochastically, conserved: $\bm{s}^2=\mathrm{const}$ and $\bm{S}^2=\mathrm{const}'$, for any realization of the noise. 

Thus if we consider the norm of the spins to be initially equal to 1 ($\mathrm{const}=1=\mathrm{const}'$), then we can cast the expression for the heavy-spin into the more convenient Landau-Lifshitz form
\begin{equation}\label{stocheqsinter}
\left\{
	\begin{array}{lcl}
	\displaystyle
	\dot{\bm{s}} &=& (\bm{\omega}+J\bm{S})\times\bm{s} \\
	\displaystyle
	\dot{\bm{S}} &=&  \displaystyle \frac{1}{1+\alpha^2}\left(\bm{\Omega}_{\textrm{eff}}-\alpha\bm{\Omega}_{\textrm{eff}}\times \bm{S}\right)\times\bm{S}
	\end{array}
	\right.
\end{equation}
where, as shown by Bertotti et al. \cite{bertotti_nonlinear_2009}, one can omit the noise in the effective field $\bm{\Omega}_{\textrm{eff}}$ for the damping term, where
\begin{equation}
	\bm{\Omega}_{\textrm{eff}}=\bm{\Omega}+J{\bm s}+\bm{\eta}
\end{equation}
Hence we have
\begin{equation}\label{stocheqs}
\left\{
\begin{array}{lcl}
\displaystyle
\dot{\bm{s}} &=& (\bm{\omega}+J\bm{S})\times\bm{s} \\
\displaystyle
\dot{\bm{S}} &=& \displaystyle \frac{1}{1+\alpha^2}\left[\bm{\Omega}+J{\bm s}+\bm{\eta}-\alpha\left(\bm{\Omega}+J{\bm s}\right)\times \bm{S}\right]\times\bm{S}
\end{array}
\right.
\end{equation}

By taking an average over the noise distribution, the equations for the first moments, that express Ehrenfest's theorem, are thus given by
\begin{equation}
\left\{
\begin{array}{rcl}
\displaystyle
\frac{d}{dt}\langle\bm{s}\rangle&=& \bigg\langle(\bm{\omega}+J\bm{S})\times\bm{s}\bigg\rangle \\
\displaystyle
\frac{d}{dt}\langle\bm{S}\rangle &=& \displaystyle \frac{1}{1+\alpha^2}\bigg\langle\left[\bm{\Omega}+J{\bm s}+\bm{\eta}-\alpha\left(\bm{\Omega}+J{\bm s}\right)\times \bm{S}\right]\times\bm{S}\bigg\rangle
\end{array}
\right.
\label{Full_HL_sys}
\end{equation}
These equations are not, yet, self--consistent: they depend on higher order correlation functions of noise and spin that have to be spelled out.
An important condition is that we want our model to describe longitudinal damping.
This means that a direct ``mean field" approximation, where $\langle \mathcal{O}_1\mathcal{O}_2\rangle=\langle\mathcal{O}_1\rangle\times\langle\mathcal{O}_2\rangle$, is not appropriate, since the equation for $\bm{s}$ would preserve its precessional character and would not generate any damping for $\bm{s}\cdot\bm{\omega}$.

To simulate such damping, we use the following {\em Ansatz} for approach to equilibrium of the average $\langle\bm{S}\times\bm{s}\rangle$: First,  we write down an evolution equation for it:
\begin{equation}
\label{eomScross_s}
\begin{array}{l}
\displaystyle
\frac{d}{dt}\langle\bm{S}\times\bm{s}\rangle\equiv
\left\langle\dot{\bm{S}}\times\bm{s} + \bm{S}\times\dot{\bm{s}}\right\rangle \\
=
\displaystyle
\bigg\langle   \left[ \bm{\Omega}+J{\bm s}+\bm{\eta}-\alpha\left(\bm{\Omega}+J{\bm s}\right)\times \bm{S} \right]\times\bm{s}\bigg\rangle \\
+ \bigg\langle\bm{S}\times \left[  (\bm{\omega}+J\bm{S})\times\bm{s} \right]
\bigg\rangle
\end{array}
\end{equation}
and, then a closure assumption for the hierarchy is proposed.

To build and close this hierarchy, the corresponding three--point cumulant of any functional of the noise is assumed to vanish, thereby leading to the expression of the higher order correlators of the spins in terms of the 1-- and 2--point functions i.e.:
\begin{equation}
\langle\langle\mathcal{F}_i[\bm{\eta}]\mathcal{G}_j[\bm{\eta}]\mathcal{H}_k[\bm{\eta}] \rangle\rangle=0.
\end{equation}
This procedure enables us to rewrite the hierarchy using only first and second order moments of the corresponding probability distributions and thus to consistently close it.
After some tedious calculations, we obtain the system for the components of the first and second-order moments for the separated and mixed distributions for $\bm{s}$ and $\bm{S}$.
As these expressions are quite cumbersome, we define the third and fourth order moments of any combination of ${\bm A},{\bm B},{\bm C},{\bm D}\in\{{\bm S},{\bm s}\}$ in the aforementioned vanishing-cumulants approximation, in order to shorten the formulae
\begin{eqnarray}
\mathcal{C}_{ijk}^{\bm{A}\bm{B}\bm{C}}&=&\langle A_iB_j\rangle\langle C_k\rangle + \langle A_iC_k\rangle\langle B_j\rangle \\
&+& \langle B_jC_k\rangle\langle A_i\rangle - 2 \langle A_i \rangle\langle B_j\rangle\langle C_k\rangle,\nonumber\\
\mathcal{E}_{ijkl}^{\bm{A}\bm{B}\bm{C}\bm{D}}&=&\langle A_iB_j\rangle\langle C_kD_l\rangle \\
&+& \langle A_iC_k\rangle\langle B_jD_l\rangle + \langle A_iD_l\rangle\langle B_jC_k\rangle \nonumber\\
&-& 2 \langle A_i \rangle\langle B_j\rangle\langle C_k\rangle \langle D_l\rangle.\nonumber
\end{eqnarray}
In this framework, we get the following set of equations of motion for $\langle \bm{s}\rangle$ and $\langle \bm{S}\rangle$.
\begin{equation}\label{effectiveSys}
\left\{
	\begin{array}{rcl}
	\displaystyle \frac{d}{dt}\langle s_i\rangle &=& \epsilon_{ijk}\left(\omega_j \langle s_k\rangle+J\langle S_j s_k\rangle\right) \\
	\displaystyle \frac{d}{dt}\langle S_i\rangle &=& \displaystyle \frac{1}{1+\alpha^2} \Bigg\{\epsilon_{ijk}\Big(\Omega_j \langle S_k\rangle+J\langle s_j S_k\rangle + \langle \eta_j S_k\rangle\Big)\Bigg\} \\
	&-&  \displaystyle \frac{\alpha}{1+\alpha^2}\bigg[\Omega_m\langle S_m S_i\rangle - \Omega_i + J\Big(\mathcal{C}_{mmi}^{\bm{s}\bm{S}\bm{S}} - \langle s_i\rangle\Big)\bigg]
	\end{array}
	\right.
\end{equation}
These equations depend on combinations of the noise ${\bm \eta}$ and of the heavy spin ${\bm S}$ at the same time.
By considering a finite value for $\tau$, we are interested in the dynamics of $\langle\eta_iS_j\rangle$.
This is  described by the Shapiro-Loginov method \cite{shapiro_formulae_1978,berdichevsky_stochastic_1999,tranchida_closing_2016} with an Ornstein-Uhlenbeck form of the noise, namely:
\begin{equation}\label{SLog}
	\frac{d}{dt}\langle\eta_iS_j\rangle=\langle\eta_i\frac{dS_j}{dt}\rangle -\frac{1}{\tau}\langle\eta_iS_j\rangle
\end{equation}
Furthermore, $\displaystyle \frac{dS_i}{dt}$ has to be expanded with its stochastic form so as to display explicitly its dependence in $\tau$, which appears through the product of two of the noise components.
As expected, this leads to quite cumbersome expressions, that can be reorganized interestingly by multiplying equation (\ref{SLog}) by $\tau$.
If the white-noise limit is what we are interested in, we may try to take the limit $\tau\to 0$ in these expressions, directly, if we assume that $\tau\frac{d}{dt}\langle\eta_iS_j\rangle$ does vanish, as $\tau\to 0$. 
The validity of this assumption is, of course, by no means obvious, but it can be checked, {\em a posteriori}, by checking the consistency of the numerical results. 

For $\tau$ multiplies the derivative terms in the differential equations and, therefore, the direct limit transforms the differential equation into an algebraic relation, a constraint. 
If the dynamics leads to an attractor, such a constraint can be interpreted as an identity between correlation functions on the attractor. 
If no such attractor exists, then there exist ``hidden conservation laws'' that lead to divergences as $\tau\to 0$, that must be treated in a way similar to Hamiltonian chaos, where one direction is expanding exponentially, while the conjugate one is contracting at the opposite rate, in order that the volume remain constant. 
The detailed study will be presented in future work--here we wish to test the consistency of this {\em Ansatz}, thereby assuming that an attractor exists. 

So we shall work in this limit and see what happens. 
The consequences of this approximation, in practice, are the algebraic relations themselves.  

The differential equations, that involve the mixed moments, $\langle\eta_i S_j\rangle$, become algebraic relations between them and the magnetization components, $\langle S_i\rangle$, {\em viz.}
\begin{equation}
\langle\eta_iS_j\rangle=-\frac{D}{1+\alpha^2}\epsilon_{ijk}\langle S_k\rangle.
\end{equation}
We proceed in a similar way for equations (\ref{effectiveSys}).
In the white-noise limit, we express $\langle\eta_i s_j \rangle$ in terms of moments involving $\bm{s}$ and $\bm{S}$ only by keeping equations for $\langle s_i\rangle$, $\langle S_i\rangle$, $\langle s_is_j\rangle$, $\langle s_iS_j\rangle$, $\langle S_iS_j\rangle$, etc.
Unfortunately expressions like $\langle\eta_i s_j\rangle$ remain and we need a way to explicit them.
In the white-noise limit and for a centered noise distribution we use the Furutsu-Novikov-Donsker theorem (FND) \cite{furutsu_statistical_1963} which yields
\begin{equation}\label{FND}
	\langle \eta_i (t)s_j(t)\rangle=\int dt'\,\langle \eta_i(t) \eta_l(t')\rangle\left\langle \frac{\delta s_j(t)}{\delta \eta_l(t')} \right\rangle.
\end{equation}
Now the expression for $\displaystyle \frac{\delta s_j(t)}{\delta \eta_l(t')}$ has to be derived.

There is only an implicit dependence of $\bm{s}$ on $\bm{\eta}$ through the exchange coupling with $\bm{S}$.
So we use the chain rule of functional differentiation to obtain the average functional derivative of ${\bm s}$ with the noise,
\begin{equation}
\label{FND1}
	\left\langle \frac{\delta s_i(t)}{\delta \eta_j(t')}\right\rangle=\int dt''\left\langle \frac{\delta s_i(t)}{\delta S_k(t'')} \frac{\delta S_k(t'')}{\delta \eta_j(t')}\right\rangle.
\end{equation}
The technical details pertaining to the derivation of the expression of $\displaystyle \frac{\delta S_k(t'')}{\delta\eta_j(t')}$ can be found, in considerable length, in the supplementary material to ref.\cite{tranchida_hierarchies_2018}.

It should be noted, that eq.~(\ref{FND1}) seems to be inconsistent, on dimensional grounds. 
This is, however, an illusion. 
The reason is that $\delta s_i(t)/\delta S_k(t')$, in fact, is not dimensionless--it carries the dimension of an inverse time, as can be understood from the example of $\delta s_I(t)/\delta s_J(t')=\delta_{IJ}\delta(t-t')$, where the dimension is ``hidden'' in the $\delta-$function; once it is integrated over, there is not any problem. 
So the correlation functions are to be understood as integrated over appropriate time intervals. 
This is the approach we follow throughout. 

Once the result of eq.~(\ref{FND1})  is inserted into equation (\ref{FND}) and the derivatives evaluated, we end up with
\begin{equation}
	\langle \eta_is_j \rangle = \frac{JD}{2(1+\alpha^2)} \big(\langle S_js_i \rangle -\delta_{ij}\langle S_ms_m\rangle\big).
\end{equation}
Combining these results in equation (\ref{effectiveSys}), we find
\begin{equation}\label{finalEffSys}
\left\{
\begin{array}{rcl}
\displaystyle \frac{d}{dt}\langle s_i\rangle &=& \epsilon_{ijk}\left(\omega_j \langle s_k\rangle+J\langle S_j s_k\rangle\right) \\
\displaystyle \frac{d}{dt}\langle S_i\rangle &=& \displaystyle \frac{1}{1+\alpha^2} \Bigg\{\epsilon_{ijk}\bigg[\Omega_j \langle S_k\rangle+J\langle s_j S_k\rangle\bigg]\Bigg\} \\
&-& \displaystyle \frac{2D}{(1+\alpha^2)^2}\langle S_i\rangle \\
&-& \displaystyle \frac{\alpha}{1+\alpha^2}\bigg[\Omega_m\langle S_m S_i\rangle - \Omega_i + J\Big(\mathcal{C}_{mmi}^{\bm{s}\bm{S}\bm{S}} - \langle s_i\rangle\Big)\bigg]
\end{array}
\right.
\end{equation}
To close this dynamical system, the evolution equations for the second order moments $\langle s_iS_j\rangle$, $\langle S_iS_j\rangle$ and $\langle s_is_j\rangle$ are required; their expressions are given in the appendix \ref{appendixA}.

These equations have to be integrated numerically and the results should be compared to the corresponding average values deduced from the sto\-chastic integration of equations (\ref{stocheqs}) in order to check the validity of the approximations we have employed.

This task is performed by using a dedicated geometric integrator \cite{hairer_geometric_2006,blanes_magnus_2009} and by directly sampling the noise.
To highlight the most striking  physical insights, we have limited our numerical study to the white-noise limit of the Ornstein-Uhlenbeck process, i.e. $\tau\to0$.
However the reader can find the framework for writing the corresponding equations for any finite value of $\tau$, in appendix \ref{appendixA}.
For a single spin, in a colored bath, the corresponding equations can be found in reference \cite{tranchida_hierarchies_2018}.

\section{Numerical integration}\label{Integration}

As mentioned above, we now, simultaneously, integrate numerically the effective expressions of the moments using equations (\ref{finalEffSys}), and the stochastic system with equation (\ref{stocheqs}) from which average values are deduced. 

Both methods have their advantages and their drawbacks.

As usual the stochastic method displays some computational and conceptual issues. 
One of the main issues is the nature of the considered noise.
To be able to perform the averaging over the different realizations of the noise, one first has to know whether or not the system is ergodic.
On top of that, the generation of a white noise relies on a numerical random process, which already is not completely trivial, so if one wishes to go further and work with noises which are not $\delta$-correlated or colored, then it becomes even more cumbersome \cite{niederreiter_random_1992}.
The computational issue is, as always, related to the fact that one has to compute a large number of realizations for the average to have some meaning. 
The more complicated the system becomes, the larger the required number of realizations is and as this number grows, simply taking more realizations into the average does no longer necessarily improve the result.

The main advantage of the effective expressions for the moments is the speed with which one can obtain ``numerically stable'' averages.

Although one also has to double check if the assumptions for the closure of the hierarchy are valid, and depending on the noise, the technicalities can be completely different, thus one needs to recompute the equations almost every time.
Another issue is due to the structure of the integration scheme itself.
The stochastic equations, in our case, enjoy non--trivial conservation laws, which are not, necessarily, inherited by  the effective system of moments anymore.

Thus although one can integrate each realization stochastically using a geometric integrator, this is no longer possible with the effective model; indeed, the norm of the first-order moment is no longer conserved: Longitudinal damping is generated! 

This means that in the long run, the stochastic solution is more stable, under the condition that the number of realizations grows with the integration time.

The numerical methods for ordinary differential equations may present some stability issues \cite{butcher_numerical_2016}.
The effective model Eq.(\ref{finalEffSys}) is integrated with 3 different numerical algorithms, viz the Rosenbrock scheme 
for stiff integration \cite{rosenbrock_general_1963}, the 7-8-th Runge-Kutta scheme \cite{verner_explicit_1978} and the 
explicit Euler scheme \cite{butcher_numerical_2016}.
We observe numerical instabilities that grow much faster with the Runge-Kutta and Euler algorithms than with the 
Rosenbrock scheme, which is not very surprising.
This is why we focus on the data provided by the last scheme only.

On the other hand, the stochastic system eq.(\ref{stocheqs}) is integrated using a $2^{\textrm{nd}}$-order Suzuki-Trotter algorithm, {\em viz.}

\begin{equation}
({\bm s}(t),{\bm S}(t))=\exp({t\cal{L}})({\bm s}(0),{\bm S}(0)) \textrm{ such that } {\cal L}={\cal L}_s+{\cal L}_S
\end{equation}
where the operators ${\cal L}_s$ and ${\cal L}_S$ are found in \cite{thibaudeau_thermostatting_2012} and the approximation is given by
\begin{equation}
e^{{t\cal{L}}}=e^{{t({\cal L}_s+{\cal L}_S)}}=e^{{\frac{t}{2}{\cal L}_s}}e^{t{\cal{L}}_S}e^{{\frac{t}{2}{\cal L}_s}}+o(\tau^3).
\end{equation}
This kind of scheme is known not only to be of high order precision, but also to preserve the structure and the phase-space evolution of the equations, so as to minimize the propagated error and thus build an integrator which one can use for longer times as well \cite{hairer_geometric_2006}. 
Moreover, for a more thorough investigation, a large set of random initial conditions on the unit sphere (for both the heavy and light spins), have been generated and propagated using a bash script to automatically generate data and check the influence of initial conditions on the long term behavior.

\begin{figure}[htp]
	\centering
	\includegraphics[width=0.7\textwidth]{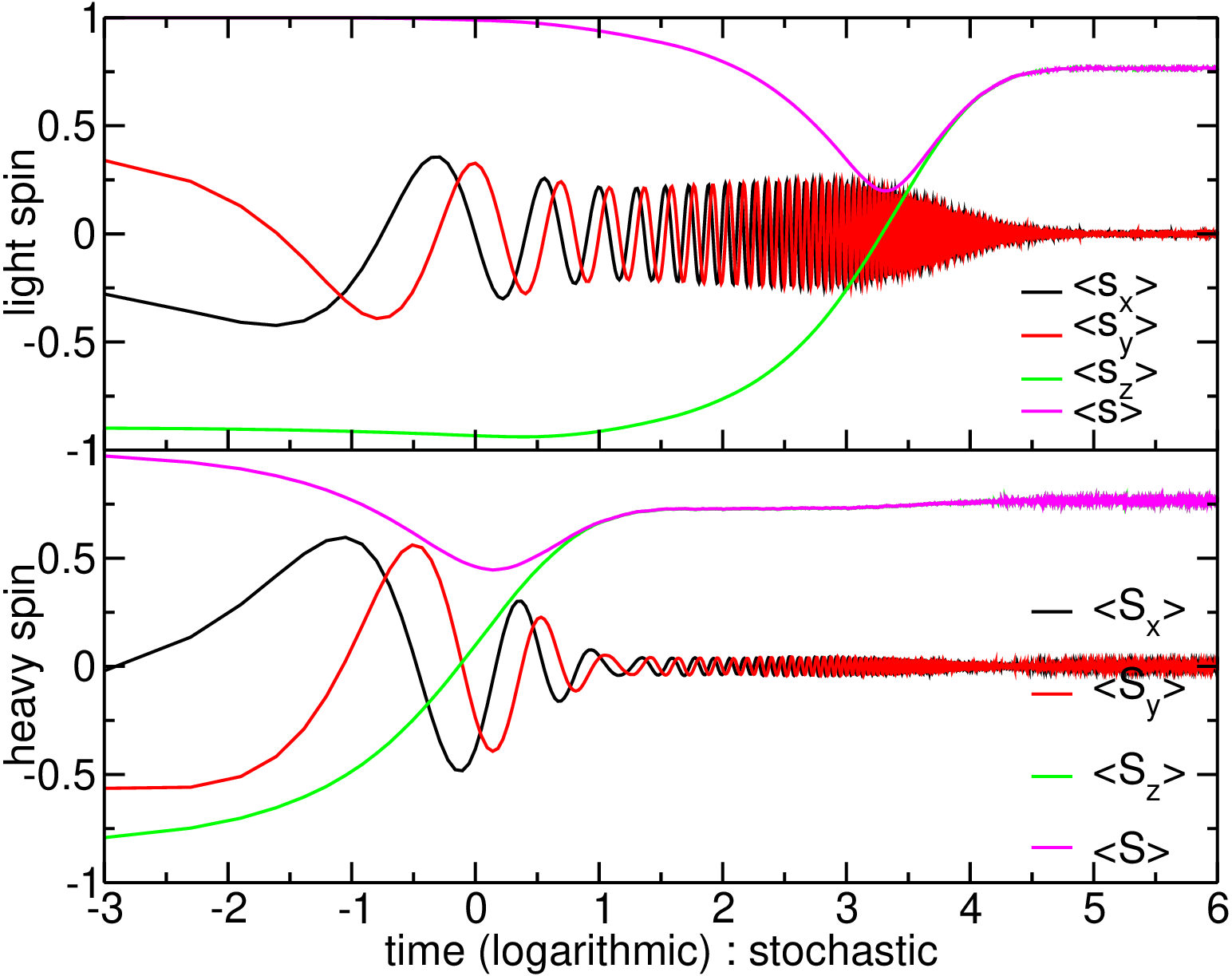}
	\includegraphics[width=0.7\textwidth]{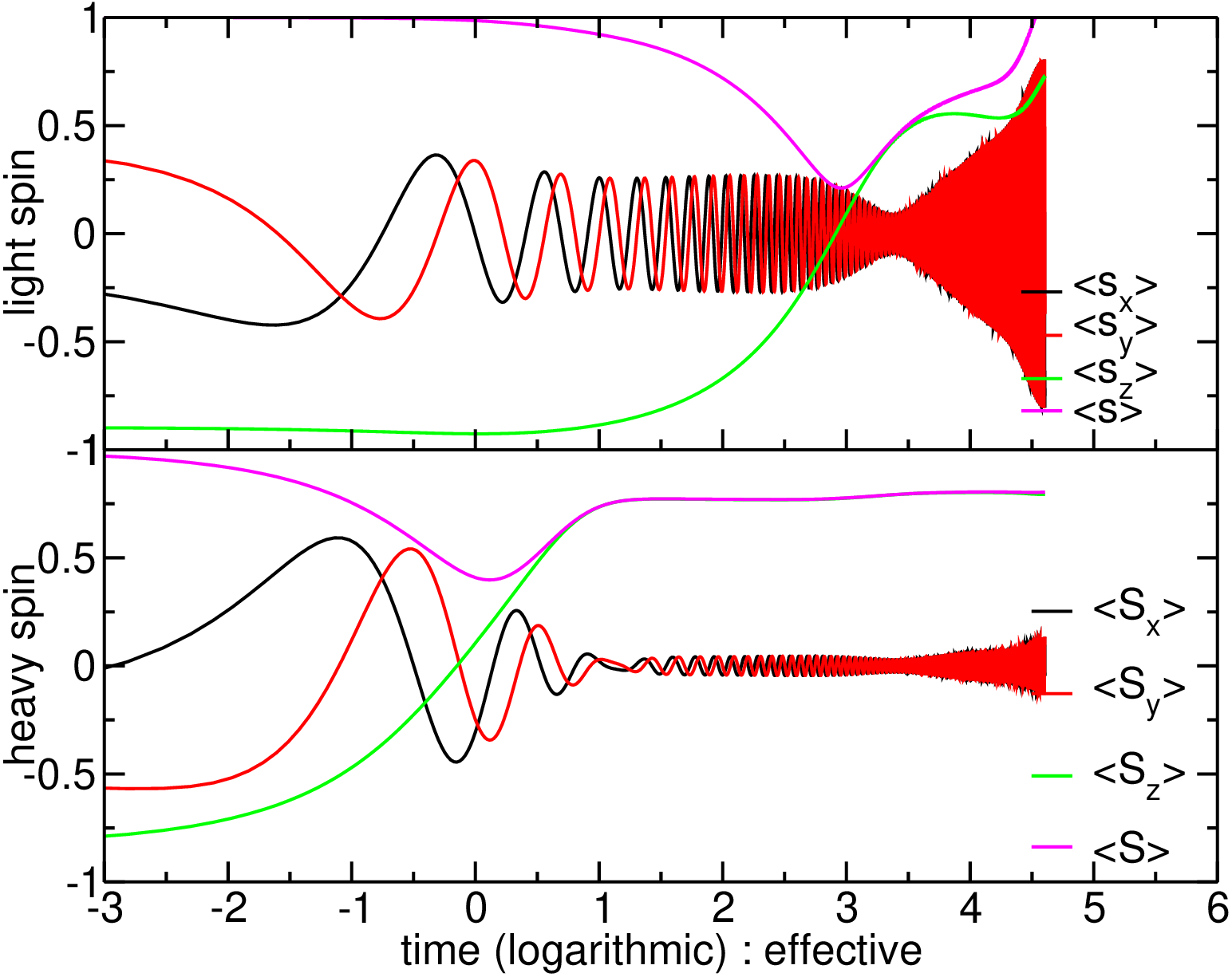}
	\caption{On the upper set, stochastic solutions ($10^5$ realizations of the noise) with the conditions : \{$s_x(0)=-0.172098$, $s_y(0)=0.409099$, $s_z(0)=-0.896114$, $S_x(0)=-0.165619$, $S_y(0)=-0.528101$, $S_z(0)=-0.832874$, $D=0.3$ GHz, $\alpha=0.2$, $\displaystyle \bm{\omega}=2\pi\bm{z}$ GHz, $\bm{\Omega}=2\pi\bm{z}$ GHz and $J=0.3$ GHz\}, On the lower set, effective solutions with the same conditions than on the upper one with \{$\langle s_is_j\rangle(0)=s_i(0)s_j(0)$, $\langle s_iS_j\rangle (0)=s_i(0)S_j(0)$, $\langle S_iS_j\rangle(0)=S_i(0)S_j(0)$\}. For numerical schemes, see text.}
	\label{IntFig}
\end{figure}

We begin by a studying a ``softly-damped" ($\alpha=0.2$) system namely with a small noise amplitude (i.e a small ``temperature") with $D=0.3$ GHz. 
Results for both the stochastic and deterministic systems are displayed in Fig.~\ref{IntFig}.

One notices that, although only the heavy-spin is explicitly coupled to the noise, the averages (first order moment) for both the heavy and light spins display longitudinal damping. 
For the stochastic model, we can see that an equilibrium solution seems to be reached, in the long run, with $||\langle\bm{s}\rangle(\infty)|| < ||\langle\bm{s}\rangle(0)||$--as is to be expected.
What should be stressed is that the dynamics does take place on the unit sphere for all times. 

Although for short times, both the effective and the stochastic model have very similar behavior, in the long run, we notice that the effective model diverges strongly enough for the integrator to ``break down'', and hence the two systems seem to become radically different. 
Of course our effective model is a first step and needs to be revised, as is discussed below, but the difference in computation time, for obtaining the time evolution, between the two approaches is quite impressive: The stochastic curve for $10^5$ realizations was produced in little over 10 hours, whereas the effective curve was produced in around a minute, on the same computer.

\begin{figure}[htp]
	\centering
	\includegraphics[width=0.7\textwidth]{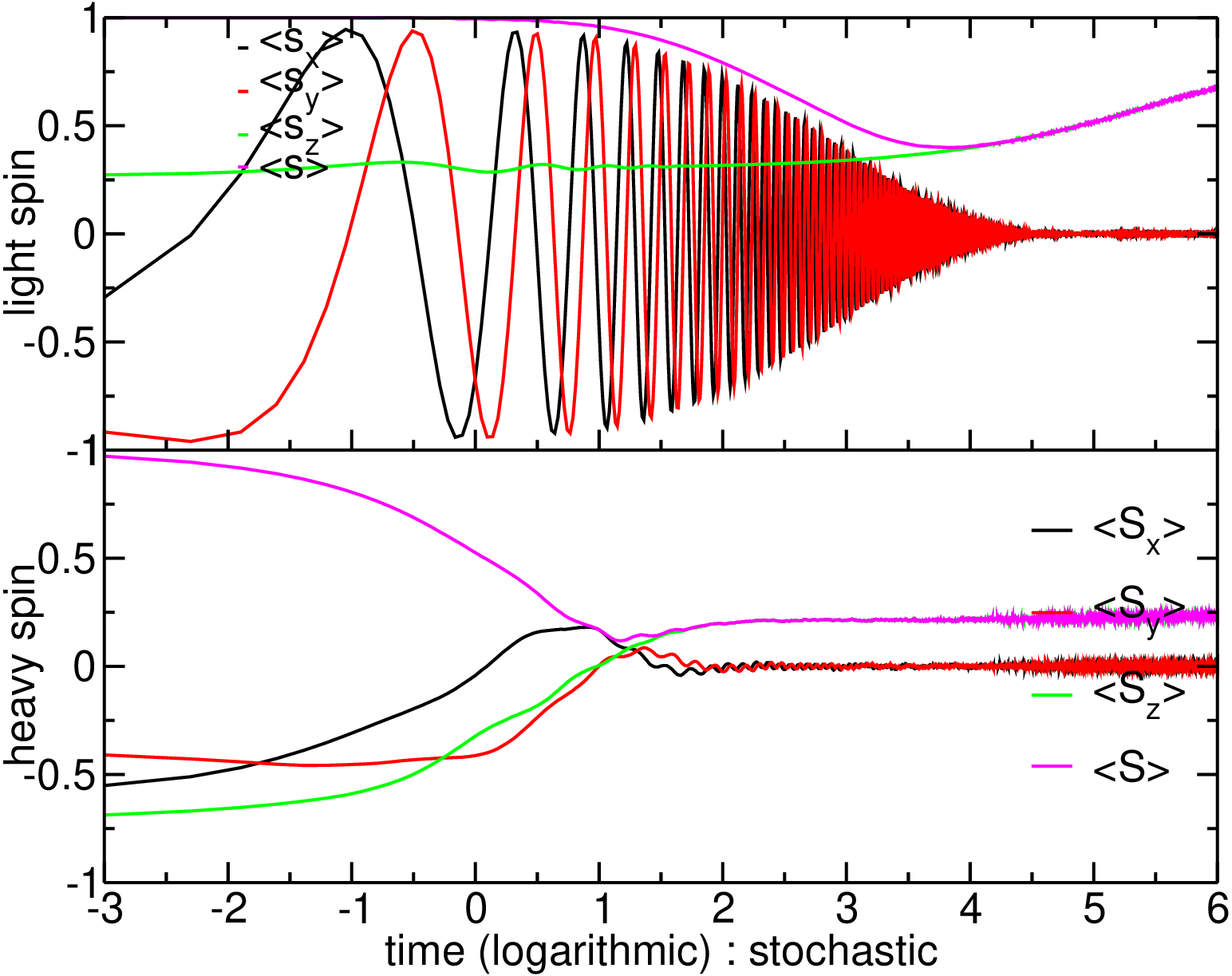}
	\includegraphics[width=0.7\textwidth]{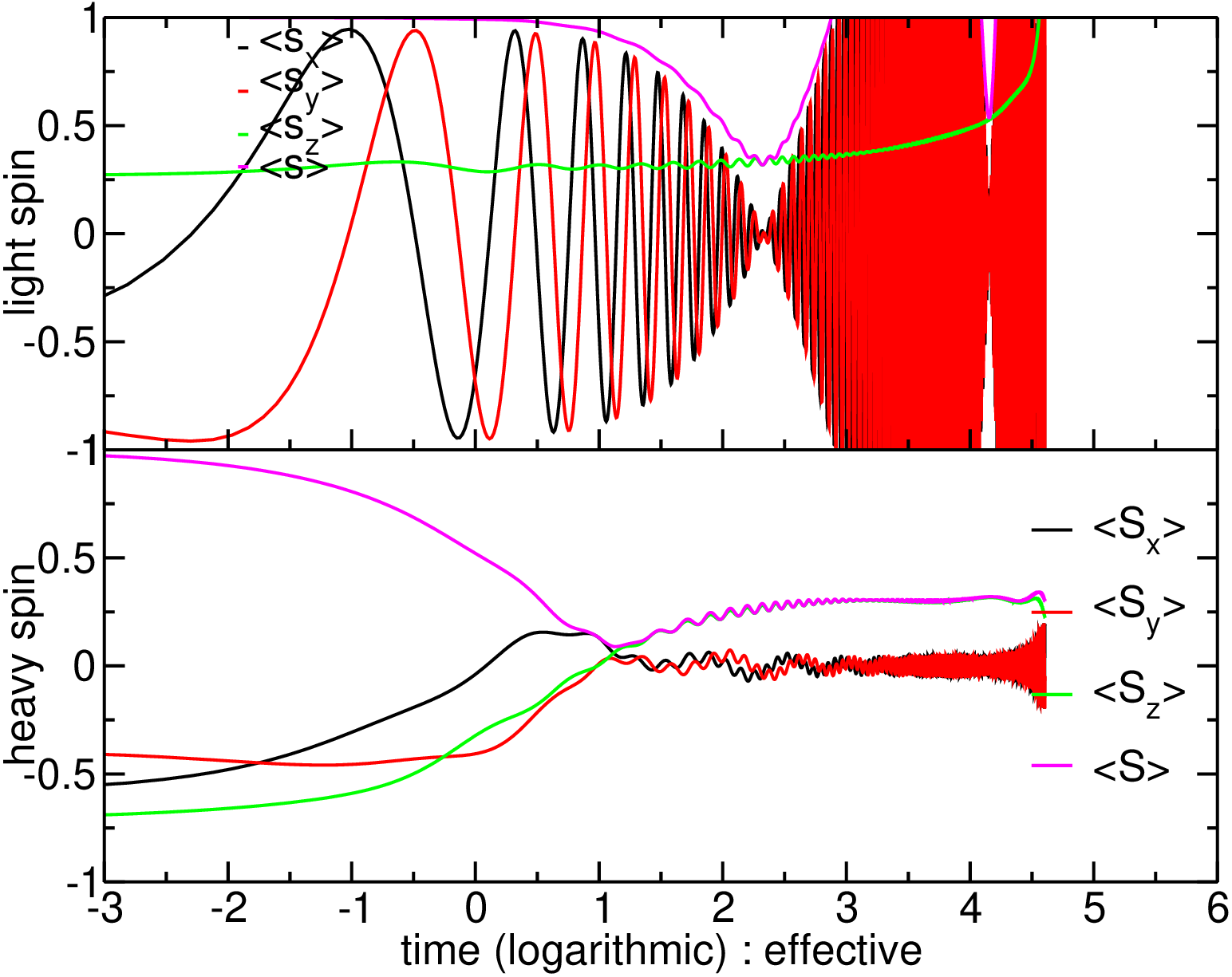}
	\caption{On the upper set, stochastic solutions ($10^5$ realizations of the noise) with the conditions \{$s_x(0)=-0.551323$, $s_y(0)=-0.790046$, $s_z(0)=0.268087$, $S_x(0)=-0.589254$, $S_y(0)=-0.385068$, $S_z(0)=-0.710283$, $D=0.3$ GHz, $\alpha=0.2$, $\displaystyle \bm{\omega}=2\pi\bm{z}$ GHz, $\bm{\Omega}={\displaystyle \frac{2\pi}{7}}\bm{z}$ GHz and $J=0.3$ GHz\},
		On the lower set, effective solutions with the same conditions than on the upper one with
		\{$\langle s_is_j\rangle(0)=s_i(0)s_j(0)$, $\langle s_iS_j\rangle (0)=s_i(0)S_j(0)$, $\langle S_iS_j\rangle(0)=S_i(0)S_j(0)$\}}
	\label{BigJFig}
\end{figure}

We, also, study the case of two different external fields for the heavy and light spins, respectively $||\bm{\omega}||=2\pi$ GHz and $||\bm{\Omega}||=\frac{2\pi}{7}$ GHz. 
Results are displayed in Fig.~\ref{BigJFig}. 

The overall dynamics remains very similar to the previous study with however a noticeable and important difference which is that the light spin does not seem to reach equilibrium anymore, within the stochastic approach. 
We also remark that as the heavy spin performs a motion of precession towards an equilibrium state, almost aligned with its external field $\bm{\Omega}$, it precesses at the frequency of $\bm{\Omega}$, and after a short transient period, its frequency shifts to $\bm{\omega}$.

\begin{figure}[htp]
	\centering
	\includegraphics[width=0.7\textwidth]{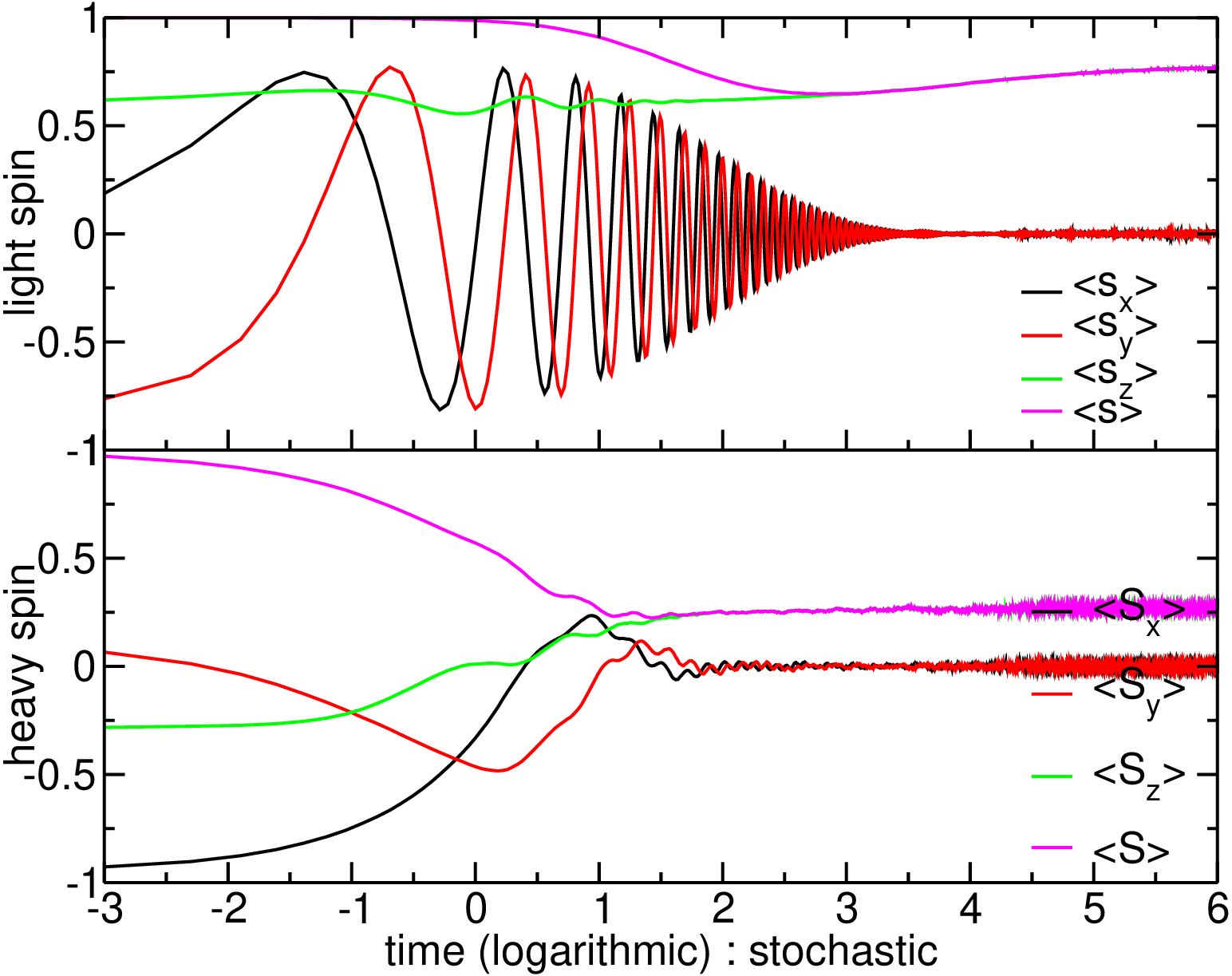}
	\includegraphics[width=0.7\textwidth]{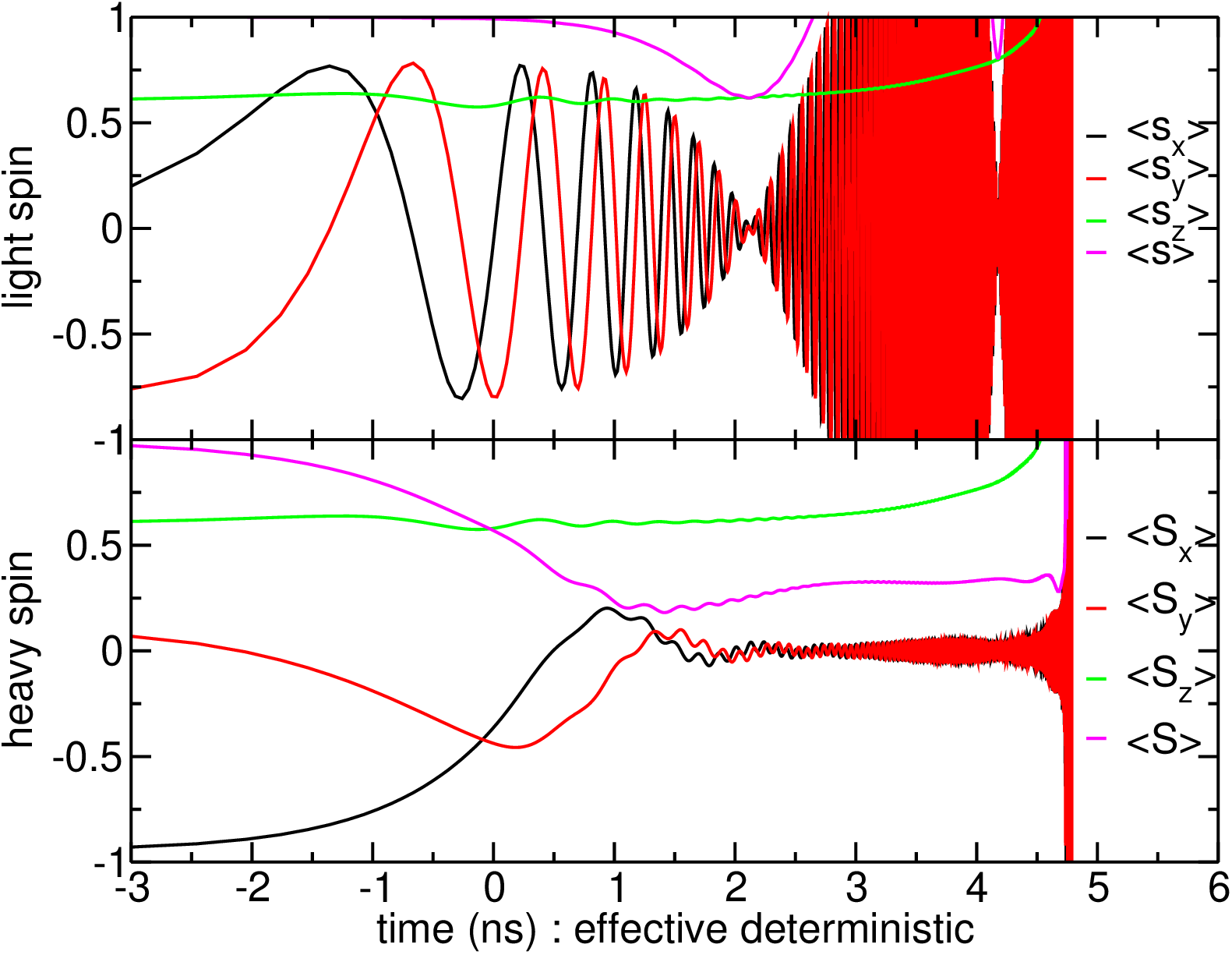}
	\includegraphics[width=0.7\textwidth]{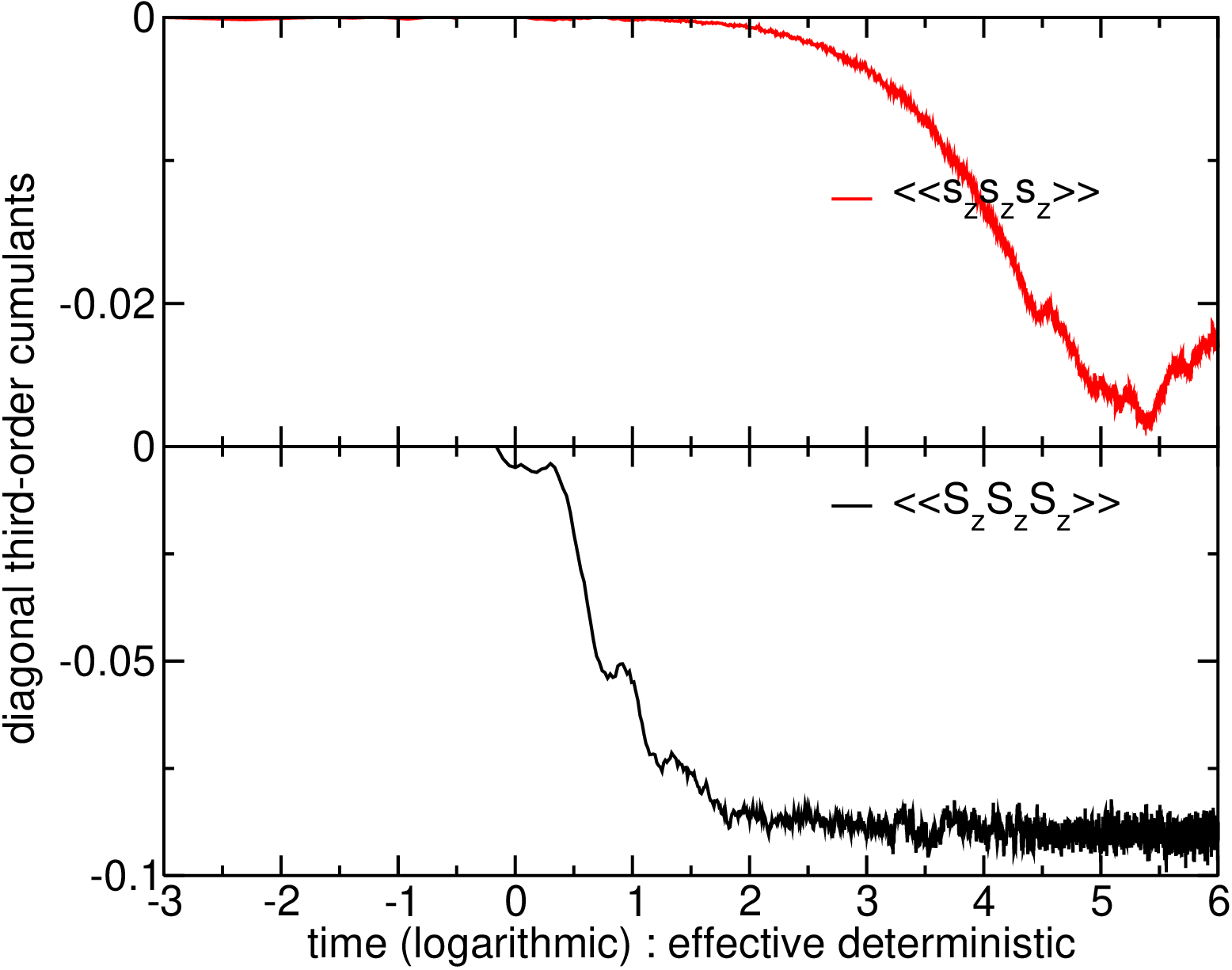}
	\caption{On the upper set, stochastic solutions ($12000$ realizations of the noise) with the conditions: \{$s_x(0)=-0.054083$, $s_y(0)=-0.797312$, $s_z(0)=0.601140$, $S_x(0)=-0.951202$, $S_y(0)=0.124420$, $S_z(0)=-0.282373$, $D=0.3$ GHz, $\alpha=0.2$, $\displaystyle \bm{\omega}=2\pi\bm{z}$ GHz, $\bm{\Omega}=\displaystyle\frac{2\pi}{7}\bm{z}$ GHz and $J=0.5$ GHz\}. On the middle set, effective solutions with the same conditions than on the upper one with \{$\langle s_is_j\rangle(0)=s_i(0)s_j(0)$, $\langle s_iS_j\rangle (0)=s_i(0)S_j(0)$, $\langle S_iS_j\rangle(0)=S_i(0)S_j(0)$\}. On the bottom set, diagonal third order cumulants for the light and heavy spin distributions are displayed.}
	\label{CumFig}
\end{figure}

Finally, we wish to report on the case, where we have imposed a stronger correlation between both systems, through a larger value of the exchange constant $J=0.5$ GHz and with the same fields $\bm{\omega}$ and $\bm{\Omega}$ as in Fig.~\ref{BigJFig}. 
The results are plotted in Fig.~\ref{CumFig} where in addition to the stochastic and effective curves, we plot the third order cumulants along the $\bm{z}$-axis for the light and heavy spins distributions.

Here we can see that for the same fields the equilibrium solution is different from Fig.~\ref{BigJFig} as we recover the convergence towards equilibrium. 
What is of even more interest is the explicit computation of the third order cumulants which, as an initial assumption, were supposed to be vanishing. 
We can see that as this is not the case, the effective model moves away from the stochastic one, once this cumulant grows significantly. 
This is interesting as it indicates that the distributions for the heavy and light spins, at least for longer times, are not distributed according to a Gaussian law. 
Further proof of this could be obtained by studying
\begin{equation}
	\Sigma_{IJ}=\langle(X_I-\langle X_I\rangle)(X_J-\langle X_J\rangle)\rangle
\end{equation}
where $X_I=\{\bm{s},\bm{S}\}$ is a ``super-vector'', comprising both the heavy and light spin components. 
If this matrix is indeed not positive definite, then the fact that our effective integration scheme diverges would actually be good news, because the system itself would be leaving an unstable equilibrium.  
This however requires a significant effort and shall be studied thoroughly elsewhere.

Overall what we expected to see, and what we wanted our model to explain, was the heavy spin more or less quickly becoming aligned with the external field, and the light spin relaxing towards an equilibrium with constant norm (lower than initially). 
On the one hand, our approach is consistent with this expectation. 
However, this does not occur without any qualification--which provides hints that the Gaussian closure is not universally valid.
This is consistent with expectations. 

For the case where both effective precession fields $\bm{\omega}$ and $\bm{\Omega}$ are the same, stochastically Fig.~\ref{IntFig} seems to display this behavior but if they are different (Fig.~\ref{BigJFig}), then the norm does not seem to converge to a particular value even for long times (i.e. 500 ns). 
As our main idea is to mimic elastic coupling, we would have expected to obtain similar equilibrium solutions as in \cite{nussle_coupling_2019} where an equilibrium is reached, but where the expectation value of the spin does not necessarily have to be conserved either. 
Depending on whether the mechanical or the magnetic system is the one with "too much" energy, there seemed to be a transfer to the other system ending up in either a transfer to the spin system (i.e. expectation value of the spin greater  than 1, negative deformations) or to the mechanical system (positive deformations and expectation value of the spin less than 1).

Furthermore, the integration scheme for the effective model breaks down at ``long'' times ($\approx$ 100 ns) and the behavior becomes already very different than the stochastic one around 40 ns.

Several reasons can be the cause of this behavior, as an example, initial conditions on the moments, since these define the distribution. 
We have assumed a Gaussian distribution--however, if the dynamics is not Gaussian, one way this can become obvious is that the covariance matrix, evaluated at the average magnetizations, is not positive--definite. 
This is hinted at by our results regarding the third order cumulant, as on can see in Fig.~\ref{CumFig}. 
Indeed it is reassuring that our procedure can probe how the Gaussian distribution, assumed in the vicinity of the average magnetization, crosses over to a non--Gaussian distribution at long times, when exploring regions ``far'' from the expectation values of the spin components. 
This deserves a more detailed study, whose difficulties, however, should be kept in mind~\cite{tranchida_closing_2016}.

This is another conceptual issue, namely, what are the motivations for the closure assumptions, which are better and in which situation; as of today, the answer is mostly empirical, trying several, and taking the ``best" ones, or worse, the ones which are not as bad as the others. 
For example if we consider stronger exchange coupling, we can definitely expect that there should be higher order correlations between the moments of the light and heavy spins probability distributions. 
This can be noticed in Fig.~\ref{CumFig}. 
It seems unavoidable that one requires to optimize the closure assumptions depending on each considered case.

A useful guide is provided in ref.~\cite{nicolis_closing_1998}, where the closure assumptions for Hamiltonian, as well as non--Hamiltonian chaotic systems were discussed.  

But still, the results are promising, and further studies are ongoing, where the vector-noise ${\bm\eta}$ is replaced by a tensorial noise, making the mechanical coupling more manifest~\cite{nussle_coupling_2019}. 
As of yet, this coupling has been taken into account through an effective damping and fluctuation model on the heavy-spin only.
However, the fact that we can find longitudinal damping, on top of the transverse damping, already, is encouraging and shows that one can probably simulate a magnetomechanically coupled model, in a way similar to the one proposed here.

\section{Conclusions and outlook}\label{Conclusions}
We have studied the case of two spins, in contact with a bath, that describes an elastic medium, through the color of the noise, and have focused on the dynamics of one, when the other can be, effectively, integrated out and defines an effective spin bath, itself.

We have solved the stochastic differential equations for the spins numerically, by directly sampling the noise and have reconstructed the 1-- and 2--point functions of one of the spins, namely the ``light"-one.

The stochastic differential equations, that define the spin dynamics, also, implicitly, define the measure over the spin configurations.
Using a truncation Ansatz, we have derived the deterministic, ordinary differential equations for the 1-- and 2--point functions of the light spin and have compared the results to those deduced from solving the stochastic equations directly.
Good agreement has been found, that indicates that the truncation assumption does capture useful properties of the dynamics.

The use of the Shapiro-Loginov theorem leads to a hierarchy of deterministic equations for the moments in a very straightforward fashion.
As expected, when the noise acts on any subset of the spin system, the latter is described by stochastic variables, defined by a non trivial probability law.
Because of the exchange, other coupled spins also become random variables, even though the noise does not directly act on them.
In the simplest aspect explored in this paper, we demonstrate that the effect of the dynamics of the noise on such a spin is given by a mixed 2--point function with an amplitude equal to the exchange constant, in a torque expression that is not the product of the average of each spin.

This leads to questions pertaining to the deterministic equations themselves--namely, whether, through their non--linearity, they can describe the transition from regular motion to deterministic chaos and, whether this deterministic chaos reflects, indeed, properties of the true dynamics.
An example for which this is the case was, indeed, found for Bloch-Bloembergen noise in ref.~\cite{tranchida_quantum_2015}, which was found to be described by the Lorenz equations. 
The corresponding study of the moment equations will be reported in future work.

A related question is, whether the random dynamics of the bath describes, in fact ``mole\-cular chaos'', as is the case in thermodynamics, where the correlation functions, with respect to the bath, are self-averaging, or deterministic chaos, where a small number of ``effective'' degrees of freedom describe the dynamics through a fractal attractor and the correlation functions are not self--averaging.

It would be interesting to explore the relation with the work in ref.~\cite{kisker_off-equilibrium_1996}, where the approach to equilibrium in spin models has been studied and with the work in refs.~\cite{ney-nifle_chaos_1993} where the notion of chaos in spin glass models has been discussed. 
The distinction between molecular and deterministic chaos, however, has not been clarified.

There is considerable activity in studying the different ways magnetic, mechanical as well as electrical properties of materials can be coupled~\cite{smejkal_electric_2017,shen_theory_2018,streib_damping_2018}; these are of particular relevance, both for understanding novel ordering scenaria, that occur, for instance, in topological anti-ferromagnets \cite{smejkal_topological_2018}, as well as in more conventional garnets such as YIG~\cite{maehrlein_dissecting_2018}.
Quite diverse experimental approaches lead to a broad range of data, that probe, in particular, the effects of the coupling to multiple baths.
This leads to new ways of constraining theoretical models and unveils new ways of understanding the physics these materials can describe, but it all seems to indicate that our knowledge of these phenomena is still lacking a proper ``global" understanding \cite{temnov_towards_2016}.
In fact, the search for experimental methods which try and isolate the excited sub-systems clearly indicates that once too many of these effects are coupled, we mainly lose the ability to correctly explain what actually happens.
So our approach towards a quantitative description of the coupling between mechanical and magnetic degrees of freedom, not only from a static but also a dynamical point of view, seems very topical, regarding the current issues in magnetism and magnetic materials more generally.

Recently a new direction has appeared, involving the control of magnetic properties through electric fields--and vice versa--exploiting the ``anomalous'' response of materials. 
In our study, the magneto--elastic coupling, in fact, is controlled by appropriate electric fields~\cite{piliposyan_internal_2014,sohn_deterministic_2017}, that, therefore, can provide the sources for the anomalous magnetic response. 
On the other hand, the effects of the ``heavy spin'', insofar as they do involve parity--breaking contributions, can be used as a proxy for probing such effects~\cite{chernodub_chiral_2017,arjona_rotational_2018}.

\appendix

\section{Closed hierarchy : Second-order moments}\label{appendixA}

Here we display the (deterministic) equations of the dynamical system for the second-order moments in the white noise limit and assuming Gaussian closure (see text):

\begin{equation}\label{finalEffSysComplete}
\begin{array}{rcl}
\displaystyle \frac{d}{dt}\langle s_i s_m\rangle &=& \epsilon_{ijk}\Bigg\{\omega_j\langle s_ks_m \rangle + J\mathcal{C}_{jkm}^{\bm{S}\bm{s}\bm{s}}\Bigg\} + \epsilon_{mjk}\Bigg\{\omega_j\langle s_ks_i \rangle \\
&+& J\mathcal{C}_{jki}^{\bm{S}\bm{s}\bm{s}}\Bigg\}\\
\displaystyle \frac{d}{dt}\langle s_iS_m\rangle &=&  \displaystyle \epsilon_{ijk}\Bigg\{\omega_j\langle s_k S_m\rangle + J \mathcal{C}_{jkm}^{\bm{S}\bm{s}\bm{S}}\Bigg\} + \frac{\epsilon_{mlp}}{1+\alpha^2}\Bigg\{\Omega_l\langle s_i S_p\rangle \\
+ J \mathcal{C}_{ilp}^{\bm{s}\bm{s}\bm{S}} &+& \displaystyle \frac{JD}{2(1+\alpha^2)}\left(\langle S_is_l\rangle-\delta_{il}\langle S_ns_n\rangle\right)\langle S_p\rangle - \frac{D}{1+\alpha^2}\epsilon_{lpn}\langle S_n\rangle\langle s_i\rangle\Bigg\} \\
&-& \displaystyle \frac{\alpha}{1+\alpha^2}\bigg[\Omega_p \mathcal{C}_{imp}^{\bm{s}\bm{S}\bm{S}} + J\mathcal{E}_{ipmp}^{\bm{s}\bm{s}\bm{S}\bm{S}} - \Omega_m\langle s_i\rangle-J\langle s_is_m\rangle\bigg] \\
\displaystyle \frac{d}{dt}\langle S_iS_m\rangle &=& \displaystyle \frac{\epsilon_{mlp}}{1+\alpha^2}\Bigg\{\Omega_l\langle S_i S_p\rangle + J \mathcal{C}_{ilp}^{\bm{S}\bm{s}\bm{S}} - \frac{D}{1+\alpha^2}\epsilon_{lin}\langle S_n\rangle\langle S_p\rangle \\
&-& \displaystyle \frac{D}{1+\alpha^2}\epsilon_{lpn}\langle S_n\rangle\langle S_i\rangle\Bigg\} + \frac{\epsilon_{ilp}}{1+\alpha^2}\Bigg\{\Omega_l\langle S_m S_p\rangle \\
&+& \displaystyle J \mathcal{C}_{mlp}^{\bm{S}\bm{s}\bm{S}} - \frac{D}{1+\alpha^2}\epsilon_{lmn}\langle S_n\rangle\langle S_p\rangle - \frac{D}{1+\alpha^2}\epsilon_{lpn}\langle S_n\rangle\langle S_m\rangle\Bigg\} \\
&-& \displaystyle \frac{\alpha}{1+\alpha^2} \Bigg\{2\Omega_p \mathcal{C}_{imp}^{\bm{S}\bm{S}\bm{S}} + 2J\mathcal{E}_{ipmp}^{\bm{S}\bm{s}\bm{S}\bm{S}} - \Omega_m\langle S_i\rangle - \Omega_i\langle S_m\rangle \\
&-& J\bigg[\langle s_iS_m\rangle + \langle s_mS_i\rangle\bigg]\Bigg\}
\end{array}
\end{equation}
What is of interest is that they, along with the equations for the first moments, $\langle s_i\rangle$, display quadratic non--linearities, so it is of interest to determine, whether they can define strange attractors, as equilibrium configurations.
\bibliographystyle{spphys}
\bibliography{cea-lmpt}
\end{document}